\documentclass[a4paper,11pt]{article}
\pdfoutput=1 % if your are submitting a pdflatex (i.e. if you have
\usepackage{jcappub} % for details on the use of the package, please
\usepackage[T1]{fontenc} % if needed

\allowdisplaybreaks

 \title{The effect of electromagnetic radiation birefringence in the field of a relativistically rotating pulsar or 
 	magnetar within the framework of vacuum nonlinear electrodynamic}

\author[a,b]{Moldir S. Seidaliyeva}
\author[a]{Victor  I. Denisov} 
\author[c,d]{Irene P. Denisova}

\affiliation[a]{Department of Physics, Moscow State University, 
	119991, Moscow,  Russia}
\affiliation[b]{Al-Farabi Kazakh National University, Al-Farabi Avenue, 71, Almaty, Kazakhstan }
\affiliation[c]{Bauman  Moscow State  Technical University, ul. Baumanskaya 2-ya, 5/1,  105005, Moscow, Russia}
\affiliation[d]{The State University of Management, 99 Ryazansky Prospekt,  109542, Moscow, Russia}

\emailAdd{seydalieva.moldyr@gmail.com}
\emailAdd{vid.msu@yandex.ru}
\emailAdd{pm.mati813@gmail.com}

\abstract{
	Within the framework of the parameterized post-Maxwellian vacuum electrodynamics, the propagation of an X-ray or 
	gamma-ray pulse 
	through the electromagnetic field of a relativistically rotating pulsar is studied.  
	
	Expressions are obtained for the trajectory of this pulse and the law The effect of electromagnetic radiation birefringence in the
	field of a relativistically rotating pulsar or magnetar within the framework of vacuum nonlinear electrodynamicsof its motion from 
	the point   ${\bf r}_s=\{x_s,y_s,z_s\},$ 
	where an X-ray or gamma-ray burst occurs at time $t=t_s$ to the point ${\bf r}_d=\{x_s,y_s,z_d\},$ where the detector of this 
	radiation is located.                                                                                  
	
	In the case when the post-Maxwellian parameters of the theory differ, $\eta_1\neq\eta_2$, the time of nonlinear electrodynamic 
	delay of electromagnetic signals transported by different normal modes is calculated.  
	The change in the polarization state of the X-ray or gamma-ray pulse after passing through the electromagnetic field of the 
	relativistically rotating pulsar is analyzed.
}

\begin{document}%
%%%%%%%%%%%%%%%%%

%}%

\maketitle 
\flushbottom

\section{Introduction}

The concept of nonlinear electrodynamics in vacuum emerged in the early 20th century. Currently, several models 
of such electrodynamics are discussed in scientific literature. The most prominent among them are the Born-Infeld nonlinear 
electrodynamics \cite{1} and the Heisenberg-Euler electrodynamics \cite{2}, which arises from quantum electrodynamics. The 
post-Maxwellian 
Lagrangian density for these theories \cite{3,4} is given by:   
\begin{equation} \label{eq1}
 L={1\over 8\pi} \Big\{[{\bf  E}^2-{\bf  B}^2]+\xi [\eta_1({\bf  E}^2-{\bf  B}^2)^2+4\eta_2({\bf  B\   E})^2]\Big\}+
O(\xi^2B^6), 
\end{equation}
where $\xi =1/B^2_q,$ $B_q=4.41\cdot 10^{13}$ .

The values of the dimensionless post-Maxwellian parameters  
$\eta_1$ and $  \eta_2$ depend on the chosen model of vacuum nonlinear electrodynamics.

According to quantum electrodynamics \cite{2}, the parameters $\eta_1$ and $ \eta_2$ have well-defined values:
$\eta_1=\alpha /(45 \pi )=5.1\cdot 10^{-5}, \ \eta_2=7\alpha /(180 \pi )=
9.0\cdot 10^{-5}$ whereas in the Born-Infeld theory \cite{1}, they are expressed in terms of an unknown constant  $a^2:$ 
$\eta_1=\eta_2=a^2B_q^2/4.$
In Maxwell's linear electrodynamics, it is evident $\eta_1=\eta_2=0.$ 

However, to date, no reliable experimental evidence for the nonlinearity of electrodynamics in vacuum has been obtained, 
although the sensitivity of current experimental techniques has approached the level necessary for detecting vacuum nonlinear 
electrodynamic effects \cite{5}. Therefore, one of the most important tasks in theoretical and mathematical physics is a comprehensive 
analysis of existing mathematical models of vacuum nonlinear electrodynamics, the development of specific methods for integrating 
the systems of nonlinear differential equations contained in these models, and the identification of those nonlinear vacuum effects 
that are most promising for experimental detection.

Since the magnetic fields generated in terrestrial laboratories, $B\sim 10^6$ Gs, are extremely weak compared to the characteristic 
quantum electrodynamic field strength $B_q=4.41\cdot 10^{13}$ Gs, there is significant interest in studying the effects of vacuum 
nonlinear electrodynamics in the fields of astrophysical objects such as pulsars and magnetars. Pulsars \cite{6,7} 
are rotating neutron stars 
that possess strong magnetic fields on the order of $B\sim 10^{13}$ Gs. Magnetars \cite{8}, 
which are also rotating neutron stars, exhibit even 
stronger magnetic fields, reaching $B\sim 10^{15}$ Gs.
In the magnetospheres of pulsars and magnetars, bursts of electromagnetic radiation are periodically generated, with spectral maxima 
in the X-ray and gamma-ray regions. This radiation is currently being actively studied in international satellite missions such 
as \cite{9} "INTEGRAL". 

The initial calculations indicate that, from the perspective of satellite experiments, the delay effect of electromagnetic pulses 
polarized in the plane of a pulsar's magnetic dipole moment (first normal mode) compared to those polarized in the orthogonal 
plane (second normal mode) is of significant interest. This effect becomes particularly pronounced during the propagation 
of these pulses from a common source within the pulsar's magnetosphere to a detector located in Earth's orbit.

Previous studies \cite{10,11} have primarily focused on slowly rotating pulsars, where the electric field is negligible. 
However, there exist 
relativistically rotating pulsars in nature, characterized by the product of their angular rotation frequency $\Omega $  and 
radius $R_0$ approaching the speed of light, i.e., $1-\Omega R_0/c<<1.$ In such pulsars, not only is the magnetic field strong, 
but the electric field is also of comparable magnitude. Consequently, the nonlinear interaction of the electromagnetic field 
of a relativistically rotating pulsar with passing X-ray and gamma radiation is expected to be more pronounced than in the case 
of a slowly rotating pulsar.

However, due to mathematical complexities, the problem of the influence of the electromagnetic field of a relativistically 
rotating pulsar on X-ray and gamma radiation has so far been addressed in a simplified formulation. Specifically, the ray 
differential equations were averaged over the pulsar's rotation period prior to integration \cite{12,13} in order to eliminate rapidly 
oscillating terms that could not be integrated analytically. As a result, the findings of these studies were not fully 
general and may have omitted experimentally significant features that could help quantify deviations of vacuum nonlinear 
electrodynamics from classical Maxwellian electrodynamics.

The goal of this study is to solve the problem of the nonlinear influence of the electromagnetic field of a relativistically 
rotating pulsar or magnetar on X-ray and gamma radiation, specifically in cases where the calculations are performed without 
averaging the equations over the pulsar's rotation period.

This work is organized as follows. In section \ref{sec2}, we refine the problem 
formulation.
Further, in section \ref{sec3}, we obtain the differential equations for the rays in the electromagnetic field of a relativistically 
rotating pulsar and discuss the method of their solution.
In section \ref{sec4}, we integrate the differential equations for the ray passing through the emitter and receiver of X-ray and gamma 
impulses.
Further, in section \ref{sec5}, we study the law of motion of X-ray and gamma impulses along these rays. Using this law, in section \ref{sec6}, 
we calculate the nonlinear electrodynamic delay of electromagnetic signals carried by different normal modes.
We conclude in section \ref{sec7}. 

\
\section{Problem statement } \label{sec2}
Let a pulsar or magnetar of radius $R_0$ with a magnetic moment $\bf m$ rotate at an angular frequency $\Omega$ around an axis 
perpendicular to the vector $\bf m$.
We assume the rotation to be relativistic: $1-\Omega R_0/c<<1.$

We place the origin of the coordinate system at the center of the neutron star forming the pulsar or magnetar and orient the 
axes of the Cartesian coordinate system such that the rotation axis of the neutron star is aligned along the z-axis.

In this coordinate system, the vectors $\bf B$ and $\bf E$  of the electromagnetic field of the rotating pulsar or magnetar 
are represented as follows \cite{14}:
\begin{align} \label{eq2}
{\bf B}({\bf r},\tau)={3({\bf m}(\tau)\ {\bf r}){\bf r}
	-r^2{\bf m}(\tau)\over r^5}-{{ \dot{\bf m}}(\tau)\over c r^2} \nonumber \\
+{3({ \dot{\bf m}}(\tau)\ {\bf r}){\bf r}\over c r^4}
+{(\ddot{\bf m}(\tau)\ {\bf r}){\bf r}-r^2\ddot{\bf m}(\tau)\over c^2 r^3}, \nonumber \\
{\bf E}({\bf r},\tau)={[{\bf r},\dot{\bf m}(\tau)]\over c r^3}+
{[{\bf r},\ddot{\bf m}(\tau)]\over c^2 r^2},
\end{align}
where the dot above the vector denotes a derivative with respect to the retarded time $\tau=t-r/c$. 

The magnetic dipole moment of the pulsar in the problem under consideration has only two non-zero components ${\bf m}(\tau)=|{\bf m}|\{\cos(\Omega \tau), \ \sin{(\Omega \tau),\ 0}\}.$

The system of equations for the electromagnetic field in the parameterized post-Maxwellian vacuum electrodynamics \eqref{eq1} outside sources is given by: 
\begin{align} \label{eq3}
\nabla_k\ U^{mk}=0, \nonumber \\
\nabla_k\  {\cal F}_{mn}+\nabla_m\  {\cal F}_{nk}
+\nabla_n\  {\cal F}_{km}=0,
\end{align}
where $\nabla_k$ -- is the covariant derivative with respect to the coordinate $x^k$ 
in spacetime with the metric tensor $g_{ik}$ Einstein's theory of gravitation, 
${\cal F}_{(3)}^{ki}={\cal F}^{kn}g_{nm}{\cal F}^{mp}g_{pl}{\cal F}^{li}.$

The tensor $ U^{mk}$ depends nonlinearly on the electromagnetic field tensor
${\cal F}^{nk}:$
\begin{equation*}
 U^{mk}=\Big[1+\xi\big(\eta_1-2\eta_2\big){\cal F}_{(2)}
\Big]{\cal F}^{mk}+4\xi\eta_2{\cal F}^{ml}{\cal F}_{ln}{\cal F}^{nk}.
\end{equation*}

Let us assume that an electromagnetic pulse with a carrier frequency $\omega.$
propagates through the electromagnetic field of a rotating pulsar. To exclude the influence of the pulsar's magnetosphere 
on the electromagnetic waves, we will consider X-ray and gamma radiation as such waves. Due to the high frequency $\omega,$
the refractive index of the magnetosphere, given by $n=1-\beta^2/\omega^2$
is practically equal to one.
From the characteristic equations obtained from the equations of nonlinear electrodynamics of the vacuum \eqref{eq3}, it follows that 
the propagation of an electromagnetic wave in spacetime with the metric tensor $g_{ik}$ of Einstein's theory of gravity in 
an external electromagnetic field $ F_{km}$ occurs along geodesics of a certain auxiliary spacetime, whose effective metric tensor 
$G_{ik}$ depends on \cite{3,15} the polarization of the wave (birefringence). 

For one normal wave, the tensor $G_{ik}$ takes the form:
\begin{equation}\label{eq4}
G^{(1)}_{ik}=g_{ik}-4\eta_1\xi
F_{in}F_{mk}g^{nm}. 
\end{equation}
For the other normal wave, which has a polarization orthogonal to the polarization of the first wave, it differs by the 
coefficient of the second term:
\begin{equation}\label{eq5}
G^{(2)}_{ik}=g_{ik}-4\eta_2\xi F_{in}F_{mk}g^{nm}.
\end{equation}

Assume that from a certain point ${\bf r}= {\bf r}_s=\{x_s,y_s,z_s\},$ located in the vicinity of the neutron star, an X-ray or 
gamma pulse was emitted at time $t=t_s$
This pulse, after passing through the electromagnetic field of the pulsar, continues to propagate and reaches a detector located 
at the point ${\bf r}={\bf r}_d.$
Let us derive the ray equations along which, according to the parametrized post-Maxwellian electrodynamics, the X-ray and gamma 
pulses propagate from the point ${\bf r}= {\bf r}_s$ to the point ${\bf r}={\bf r}_d$, and also determine the equations of motion 
governing the propagation of electromagnetic pulses along these rays.

In the general case of an arbitrarily located X-ray and gamma-ray detector, the analytical solution to the formulated problem 
becomes exceedingly complicated. Therefore, to simplify further calculations, we choose the coordinate system such that 
${\bf r}_d,$  the detector position is given by the ${\bf r}_d=\{x_s,y_s z_d\}.$ Additionally, we assume that the condition 
$\sqrt{x^2_s+y^2_s}>R_0,$ holds, which implies that the radiation beam propagates without intersecting the interior of the neutron 
star. This assumption allows us to neglect complex interactions with the star's dense matter and focus on the influence of external 
electromagnetic and gravitational fields.

A detailed analysis demonstrates that, in this particular case, the same nonlinear effects can be identified with sufficient 
clarity as in the more general scenario where the coordinate axes are arbitrarily oriented with respect to the line connecting 
the source and the X-ray and gamma-ray detector. This configuration allows for a more tractable mathematical treatment while 
preserving the essential physical characteristics of vacuum nonlinear electrodynamics.
Neutron stars possess significant gravitational fields. However, as shown in previous studies, 
the effects arising from the influence of a pulsar's gravitational 
field on the propagation of electromagnetic waves are relatively small in magnitude and can be treated as additive corrections 
to the nonlinear electrodynamic effects. Therefore, in the present work, we do not consider these gravitational contributions 
and instead adopt the Minkowski metric tensor $g_{ik}$
as the Einsteinian spacetime metric $g_{ik}=diag\{1,-1,-1,-1\}.$

\section{Differential equations for rays in the electromagnetic field of a relativistically rotating pulsar}\label{sec3}
The differential equations for the rays of an electromagnetic wave in the problem under consideration can be derived \cite{16} from the 
equation for isotropic geodesics:
\begin{equation} \label{eq6}
{dK^n\over d\Sigma}+\Gamma^n_{mp}K^mK^p=0,
\end{equation}
where  $\Sigma$ -- a certain affine parameter, $K^m=dx^m/d\Sigma$ -- wave four-vector, $\Gamma^n_{mp}$ -- Christoffel symbols of the effective spacetime with the metric tensor $G_{nm}^{(1)}$ for normal waves of 
the first type and $G_{nm}^{(2)}$ - for normal waves of the second type.
As in the theory of gravity \cite{16}, the equations \eqref{eq6} have the first integral:
\begin{equation} \label{eq7}
G_{np}^{(1,2)}K^nK^p=0.
\end{equation}

Using the expressions \eqref{eq4} and \eqref{eq5}, we can write the components of the metric tensor $G_{np}^{(1,2)}$
through the intensities ${\bf E}({\bf r},\tau)$ and  ${\bf B}({\bf r},\tau)$
electromagnetic field \eqref{eq2} generated by a relativistically rotating pulsar:
\begin{align} \label{eq8}
G_{00}^{(1,2)}=1-4\xi\eta_{1,2}{\bf E}^2({\bf r},\tau),\ \ \ \ \  
G_{01}^{(1,2)}=4\xi\eta_{1,2}\big[E_yB_z-E_zB_y\big], \nonumber \\
G_{02}^{(1,2)}=4\xi\eta_{1,2}\big[E_zB_x-E_xB_z\big],\ \ \ \ \ 
G_{03}^{(1,2)}=4\xi\eta_{1,2}\big[E_xB_y-E_yB_x\big], \nonumber \\
G_{\alpha\beta}^{(1,2)}=-\delta_{\alpha\beta}-
4\xi\eta_{1,2}\Big\{{\bf B}^2({\bf r},\tau)
\delta_{\alpha \beta}-E_\alpha E_\beta-B_\alpha B_\beta\Big\},
\end{align}
where $\delta_{\alpha \beta}$ -- the Kronecker symbol.                                     

Following the work \cite{3}, we transition from the equations \eqref{eq6} and the first integral \eqref{eq7} of differentiation 
with respect to the 
parameter $\Sigma$  to differentiation with respect to the coordinate $z$ according to the expression $d/d\Sigma=K^3d/dz.$
As a result, we obtain:
\begin{align}\label{eq9}
{d^2 x^0\over dz^2}=-\big\{\Gamma^0_{mp}
-{d x^0\over dz}\Gamma^3_{mp}
\big\}{d x^p\over dz}{d x^m\over dz},\nonumber \\
{d^2 x\over dz^2}=-\big\{\Gamma^1_{mp}-{d x\over dz}\Gamma^3_{mp}
\big\}{d x^p\over dz}{d x^m\over dz}, \nonumber \\
{d^2 y\over dz^2}=-\big\{\Gamma^2_{mp}
-{d y\over dz}\Gamma^3_{mp}
\big\}{d x^p\over dz}{d x^m\over dz},\ \ \
G^{(1,2)}_{np}{dx^n\over dz}{dx^p\over dz}=0.
\end{align}

The solution to these equations will allow us to find the ray equations $x=x(z), y=y(z) $, and the law of motion of the X-ray or 
gamma impulses along these rays $x^0=ct=ct(z).$

The components \eqref{eq8} of the metric tensor $G_{np}^{(1,2)}$ in our problem differ from the Minkowski tensor by dimensionless terms of the form
$\xi\eta_{1,2}E_\alpha E_\beta,$  
$\xi\eta_{1,2}B_\alpha E_\beta$  $\xi\eta_{1,2}B_\alpha B_\beta.$
These terms near the neutron star $r<10R_0,$ where the main part of the nonlinear electrodynamic interaction with X-ray and gamma 
impulses takes place, are small $\sim 10^{-5}.$ Therefore, in the right-hand sides of the equations \eqref{eq9}, there is a small 
parameter $\sim 10^{-5},$ with powers of which the solution to the system of nonlinear equations \eqref{eq9} can be constructed.

Restricting ourselves to the linear terms only, we will seek the solution to the equations \eqref{eq9} in the form:
\begin{align}\label{eq10}
x^0=ct_0(z)+\xi \eta_{1,2}{\bf m}^2ct_1(z), \nonumber \\
x=x_0(z)+\xi \eta_{1,2}{\bf m}^2x_1(z),\ \ \ \ \ \ 
y=y_0(z)+\xi \eta_{1,2}{\bf m}^2y_1(z).
\end{align}
As boundary conditions for the equations \eqref{eq9}, we require that the X-ray or gamma pulse at time $t=t_s$ is located at the point 
${\bf r}= {\bf r}_s$, and the ray passes through the points ${\bf r}= {\bf r}_s$
and ${\bf r}= {\bf r}_d.$ 
Then, from the relations \eqref{eq10}, we obtain the conditions:
\begin{align} \label{eq11}
x_0(z_s)=x_0(z_d)=x_s,\ y_0(z_s)=y_0(z_d)=y_s,\ t_0(z_s)=t_s,\ \nonumber \\
ct_1(z_s)=x_1(z_s)=x_1(z_d)=y_1(z_s)=y_1(z_d)=0 .
\end{align}
Substituting the expressions \eqref{eq10} into the equations \eqref{eq9} and expanding them by the small parameter of the problem, 
in the zeroth 
approximation we will have:
\begin{align*}
{d^2ct_0(z)\over dz^2}={d^2x_0(z)\over dz^2}={d^2y_0(z)\over dz^2}=0, \\
\Big({dct_0(z)\over dz}\Big)^2-\left({dx_0(z)\over dz}\right)^2-
\left({dy_0(z)\over dz}\right)^2=1.
\end{align*}
These equations, with the conditions \eqref{eq11}, result in:
\begin{equation*}
ct_0(z)=ct_s+z-z_s,\  \ \ x_0(z)=x_s=\rho_s\cos\varphi_s,
\\\  y_0(z)=y_s=\rho_s\sin\varphi_s,
\end{equation*}
where a convenient parameterization  for further calculations, $\rho_s$, $\varphi_s$ of the coordinates $x_s$ and $y_s$  of 
the point where the X-ray or gamma-ray burst occurred, has been introduced.

In the first approximation, using the small parameter, the equations \eqref{eq9} take the form:
\begin{align} \label{eq12}
{d^2 ct_1(z)\over dz^2}=\Big[-{90\rho_s^4z\over R^{12}}
+{12\rho_s^2z\over R^{10}}(13k^2\rho_s^2+4)
-{132k^2\rho_s^4\over R^9}-\nonumber \\
-{6k^2\rho_s^2z\over R^8}(3k^2\rho_s^2+20)
+{2k^2\rho_s^2\over R^7}(19k^2\rho_s^2+54)
+{40k^4\rho_s^2z\over R^6}-\nonumber \\
-{40k^4\rho_s^2\over R^5}
\Big]\cos[2k\big(R-z\big)+2\Psi_1\big)]
-\Big[{180k\rho_s^4z\over R^{11}}-{66k\rho_s^4\over R^{10}}-\nonumber \\
-{24k\rho_s^2z\over R^9}(3k^2\rho_s^2+4)
+{6k\rho_s^2\over R^8}(17k^2\rho_s^2+9)
+{2k^3\rho_s^2z\over R^7}(k^2\rho_s^2+46)-\nonumber \\
-{2k^3\rho_s^2\over R^6}(3k^2\rho_s^2+46)
-{8k^5\rho_s^2z\over R^5}+{8k^5\rho_s^2\over R^4}\Big]
\sin[2k\big(R-z\big)+2\Psi_1\big)]
-{90\rho_s^4z\over R^{12}}-\nonumber \\
-{24\rho_s^2z\over R^{10}}(k^2\rho_s^2-2)
-{6z\over R^8}(k^4\rho_s^4+2)
+{10k^4\rho_s^4\over R^7}+{16k^4\rho_s^2z\over R^6}
-{20k^4\rho_s^2\over R^5}-{8k^4z\over R^4}+{8k^4\over R^3},\nonumber \\
{d^2 x_1(z)\over dz^2}=x_s\Big[{90\rho_s^4\over R^{12}}
-{12\rho_s^2\over R^{10}}(13k^2\rho_s^2+8)
-{72k^2\rho_s^2z\over R^9}+{6k^2\rho_s^2\over R^8}(3k^2\rho_s^2+26)+\nonumber \\
+{24k^4\rho_s^2z\over R^7}-{24k^4\rho_s^2\over R^6}\Big]
\cos[2k\big(R-z\big)+2\Psi_1\big)]
+x_s\Big[{180k\rho_s^4\over R^{11}}+{36k\rho_s^2z\over R^{10}}-\nonumber \\
-{24k\rho_s^2\over R^9}(3k^2\rho_s^2+8)
-{60k^3\rho_s^2z\over R^8}+{2k^3\rho_s^2\over R^7}(k^2\rho_s^2+38)
+{4k^5\rho_s^2z\over R^6}-{4k^5\rho_s^2\over R^5}\Big]\times\nonumber \\
\times
\sin[2k\big(R-z\big)+2\Psi_1\big)]-\Big[{60k\rho_s^3\over R^9}
+{24k\rho_sz\over R^8}-{4k\rho_s\over R^7}(8k^2\rho_s^2+15)-\nonumber \\
-{40k^3\rho_sz\over R^6}
+{40k^3\rho_s\over R^5}\Big]
\sin[2k\big(R-z\big)+\Psi_0\big)]
-\Big[{30\rho_s^3\over R^{10}}-{6\rho_s\over R^8}(11k^2\rho_s^2+5)-\nonumber \\
-{48k^2\rho_sz\over R^7}+{2k^2\rho_s\over R^6}(3k^2\rho_s^2+32)
+{12k^4\rho_sz\over R^5}-{12k^4\rho_s\over R^4}
\Big]\cos[2k\big(R-z\big)+\Psi_0\big)]+\nonumber \\
+{90\rho_s^4x_s\over R^{12}}+{6\rho_s^2x_s(4k^2\rho_s^2-21)\over R^{10}}
+{6x_s(k^4\rho_s^4-5k^2\rho_s^2+3)\over R^8}+{24ky_sz\over R^8}+\nonumber \\
+{20k^3\rho_s^2y_s\over R^7}-{2k^3(3k\rho_s^2x_s-8y_sz)\over R^6}
+{4k^3(kx_sz-4y_s)\over R^5}-{4k^4x_s\over R^4},\nonumber \\
{d^2 y_1(z)\over dz^2}=y_s\Big[{90\rho_s^4\over R^{12}}
-{12\rho_s^2\over R^{10}}(13k^2\rho_s^2+8)
-{72k^2\rho_s^2z\over R^9}+{6k^2\rho_s^2\over R^8}(3k^2\rho_s^2+26)+\nonumber \\
+{24k^4\rho_s^2z\over R^7}-{24k^4\rho_s^2\over R^6}\Big]
\cos[2k\big(R-z\big)+2\Psi_1\big)]
+y_s\Big[{180k\rho_s^4\over R^{11}}+{36k\rho_s^2z\over R^{10}}-\nonumber \\
-{24k\rho_s^2\over R^9}(3k^2\rho_s^2+8)
-{60k^3\rho_s^2z\over R^8}+{2k^3\rho_s^2\over R^7}(k^2\rho_s^2+38)
+{4k^5\rho_s^2z\over R^6}-{4k^5\rho_s^2\over R^5}\Big]\times\nonumber \\
\times
\sin[2k\big(R-z\big)+2\Psi_1\big)]-\Big[{60k\rho_s^3\over R^9}
+{24k\rho_sz\over R^8}-{4k\rho_s\over R^7}(8k^2\rho_s^2+15)-\nonumber \\
-{40k^3\rho_sz\over R^6}
+{40k^3\rho_s\over R^5}\Big]
\cos[2k\big(R-z\big)+\Psi_0\big)]
+\Big[{30\rho_s^3\over R^{10}}-{6\rho_s\over R^8}(11k^2\rho_s^2+5)-\nonumber \\
-{48k^2\rho_sz\over R^7}+{2k^2\rho_s\over R^6}(3k^2\rho_s^2+32)
+{12k^4\rho_sz\over R^5}-{12k^4\rho_s\over R^4}
\Big]\sin[2k\big(R-z\big)+\Psi_0\big)]+\nonumber \\
+{90\rho_s^4y_s\over R^{12}}+{6\rho_s^2y_s(4k^2\rho_s^2-21)\over R^{10}}
+{6y_s(k^4\rho_s^4-5k^2\rho_s^2+3)\over R^8}-{24kx_sz\over R^8}-\nonumber \\
-{20k^3\rho_s^2x_s\over R^7}-{2k^3(3k\rho_s^2y_s+8x_sz)\over R^6}
+{4k^3(ky_sz+4x_s)\over R^5}-{4k^4y_s\over R^4},
\end{align}

where the following notations have been introduced:
$R=\sqrt{z^2+\rho_s^2},\ \rho_s^2=x_s^2+y_s^2,\ 
\Psi_0=2k(z_s-ct_s)+\varphi_s,$
$\Psi_1=k(z_s-ct_s)+\varphi_s.$

The first integral in this approximation gives the relation:
\begin{align}\label{eq13}
{d ct_1(z)\over dz}=\Big[{9\rho_s^4\over R^{10}}
-{3\rho_s^2\over R^8}(5k^2\rho_s^2+2)
-{12k^2\rho_s^2z\over R^7}+{k^2\rho_s^2\over R^6}(k^2\rho_s^2+14)+\nonumber \\
+{2k^4\rho_s^2z\over R^5}-{2k^4\rho_s^2\over R^4}
\Big]\cos[2k\big(R-z\big)+2\Psi_1\big)]+\Big[{18k\rho_s^4\over R^9}
+{6k\rho_s^2z\over R^8}-\nonumber \\
-{6k\rho_s^2\over R^7}(k^2\rho_s^2+2)-{8k^3\rho_s^2z\over R^6}
+{8k^3\rho_s^2\over R^5}\Big]
\sin[2k\big(R-z\big)+2\Psi_1\big)]+{9\rho_s^4\over R^{10}}+\nonumber \\
+{3\rho_s^2\over R^8}(k^2\rho_s^2-2)+{(k^4\rho_s^4+2)\over R^6}
+{2k^4\rho_s^2z\over R^5}-{4k^4\rho_s^2\over R^4}
-{4k^4z\over R^3}+{4k^4\over R^2}.
\end{align}
It is easy to verify that the first equation of the system \eqref{eq12} is obtained by differentiating the relation \eqref{eq13} with respect to $z$. 
Therefore, instead of the first equation of the system \eqref{eq12}, we will solve equation \eqref{eq13}, as it is simpler.

Thus, to find $ct_1(z),\ x_1(z)$ and $y_1(z)$
we need to solve two second-order ordinary differential equations of the form
\begin{equation*}
{d^2W_{x,y}(z)\over dz^2}=H_{x,y}(z)
\end{equation*}
and one first-order differential equation 
\begin{equation*}
	{dct_1(z)\over dz}=H(z)
\end{equation*}
with known right-hand sides $H_{x,y}(z),\ H(z)$, and homogeneous boundary conditions
$ct_1(z_s)=x_1(z_s)=y_1(z_s)=x_1(z_d)=y_1(z_d)=0$.  
Since this problem has readily obtainable solutions 
\begin{align*}
ct_1(z)=\int\limits^{z}_{z_s} H(\sigma)d\sigma,\\
W_{x,y}(z)=z\int\limits^{z}_{z_s}H_{x,y}(\sigma)d\sigma
-\int\limits^{z}_{z_s}\sigma H_{x,y}(\sigma)d\sigma+\\
+{(z-z_s)\over (z_d-z_s)}\Big\{\int\limits^{z_d}_{z_s}\sigma H_{x,y}(\sigma)d\sigma
-z_d\int\limits^{z_d}_{z_s} H_{x,y}(\sigma)d\sigma\Big\}.
\end{align*}
And since the second powers of $\sigma$  in the integrands can be reduced using the relation $\sigma^2=R^2-\rho_s^2,$ 
the entire problem of determining the ray trajectories in the electromagnetic field of a relativistically rotating pulsar and 
the propagation laws of X-ray and gamma-ray pulses along these rays reduces to the calculation of integrals with variable upper 
limits of the form 
\begin{align}\label{eq14}
S_0(z,n,\Psi)=\int\limits^z_0 {\sin[2k(\sqrt{\sigma^2+\rho_s^2}-\sigma)+\Psi]
	\over(\sqrt{\sigma^2+\rho_s^2})^n}d\sigma,\nonumber \\
C_0(z,n,\Psi))=\int\limits^z_0 {\cos[2k(\sqrt{\sigma^2+\rho_s^2}-\sigma)+\Psi]
	\over(\sqrt{\sigma^2+\rho_s^2})^n}d\sigma, \nonumber \\
S_1(z,n,\Psi))=\int\limits^z_0 {\sigma\sin[2k(\sqrt{\sigma^2+\rho_s^2}-\sigma)+\Psi]
	\over(\sqrt{\sigma^2+\rho_s^2})^n}d\sigma, \nonumber \\
C_1(z,n,\Psi))=\int\limits^z_0 {\sigma\cos[2k(\sqrt{\sigma^2+\rho_s^2}-\sigma)+\Psi]
	\over(\sqrt{\sigma^2+\rho_s^2})^n}d\sigma.
\end{align}
Here, the integer $n$ in expressions \eqref{eq14} ranges from 12 to 4, and $\Psi$  and $\rho_s$ -- are certain constants. 

These integrals 
are not independent, as it can be shown \cite{17} that for $n>2$ and $\rho_s> 0$ 
the functions $S_0(z,n,\Psi),\ C_0(z,n,\Psi),\ S_1(z,n,\Psi)$ 
and $C_1(z,n,\Psi)$
can be expressed in terms of elementary functions and six base functions.
$S_0(z,1,\Psi), C_0(z,1,\Psi),$ 
$S_0(z,2,\Psi),\ C_0(z,2,\Psi),$ $S_1(z,2,\Psi)$ 
and $C_1(z,2,\Psi)$.

\section{Integration of the Differential Equations for the Ray passing through the Emitter and Detector of X-ray and Gamma-ray Pulses}\label{sec4}

To determine the influence of the electromagnetic field of a relativistically rotating pulsar on the ray along which X-ray and 
gamma-ray radiation propagates from the source to the detector, we consider the following differential equations with homogeneous 
boundary conditions:
\begin{align*}
{d^2x_1\over dz^2}=H_x(z),\ \ \ \ {d^2y_1\over dz^2}=H_y(z), \\
x_1(z_s)=y_1(z_s)=x_1(z_d)=y_1(z_d)=0.
\end{align*}
By solving these equations, and also considering the corresponding solutions  of the zero approximation, we obtain 
the dependence of $x(z)$ as:
\begin{equation} \label{eq15}
x(z)=x_s+\eta\xi{\bf m}^2\Big\{F_x(z)-F_x(z_s)+
{(z-z_s)\over (z_d-z_s)}\Big[F_x(z_s)-F_x(z_d)\Big]\Big\},
\end{equation}
where $\eta=\eta_1$ for the electromagnetic pulse carried by the first normal wave, and $\eta=\eta_2$ for the one carried by 
the second normal wave, the following notation has been introduced for simplicity:
\begin{align} \label{eq16}
F_x(z)=
-{9\rho_s^2x_s\over 8R^8}-{x_s(8k^2\rho_s^2-15)\over 16R^6}
-{(16k^4\rho_s^4x_s-24k^2\rho_s^2x_s+64kzy_s-57x_s)\over 64\rho_s^2R^4}
-\nonumber \\
-{4k^3y_s\over 3R^3}+{1\over 128\rho_s^4R^2}(16k^4\rho_s^4 x_s
-256k^3\rho_s^2zy_s+120k^2\rho_s^2x_s-192kzy_s+285x_s)-\nonumber \\
-{4k^4zx_s\over 3\rho_s^2R}-{1\over 128\rho_s^7}
(304k^4z\rho_s^4x_s+256k^3\rho_s^4y_s+360k^2\rho_s^2zx_s
+192k\rho_s^2y_s+\nonumber \\
+855zx_s)arctg\Big({z\over \rho_s}\Big)
-x_s\Big[{9k\rho_s^2\over 4R^7}-{k(4k^2\rho_s^2+5)\over 8R^5}
+{5kz\over 16\rho_s^2R^4}+\nonumber \\
+{k(4k^2\rho_s^2-35)\over 32\rho_s^2R^3}
+{kz(16k^2\rho_s^2+25)\over 16\rho_s^4R^2}
-{k\over 64\rho_s^4R}(16k^4\rho_s^4-68k^2\rho_s^2+175)-\nonumber \\
-{k(z+R)\over 64\rho_s^6}(144k^4\rho_s^4-64k^2\rho_s^2-525)
\Big]\sin[2k(\sqrt{z^2+\rho_s^2}-z)+2\Psi_1]-\nonumber \\
-x_s\Big[{9\rho_s^2\over 8R^8}-{(28k^2\rho_s^2+5)\over 16R^6}
-{(12k^2\rho_s^2+35)\over 64\rho_s^2R^4}
-{5k^2z\over 8\rho_s^2R^3}+\nonumber \\
+{(16k^4\rho_s^4-180k^2\rho_s^2-175)\over 128\rho_s^4R^2}
+{5k^2z\over 8\rho_s^4R}(4k^2\rho_s^2-5)+\nonumber \\
+{(64k^6\rho_s^6+580k^2\rho_s^2+525)\over 128\rho_s^6}
\Big]\cos[2k(\sqrt{z^2+\rho_s^2}-z)+2\Psi_1]+\nonumber \\
+\Big[{5k\rho_s\over 4R^5}+{3kz\over 8R^4\rho_s}
-{5k\over 16\rho_sR^3}
+{kz(6k^2\rho_s^2+7)\over 8\rho_s^3R^2}
+{k\over 32\rho_s^3R}(16k^2\rho_s^2-25)+\nonumber \\
+{k(z+R)\over 32\rho_s^5}(16k^4\rho_s^4-28k^2\rho_s^2+75)
\Big]\sin[2k(\sqrt{z^2+\rho_s^2}-z)+\Psi_0]+\nonumber \\
+\Big[{5\rho_s\over 8R^6}-{(48k^2\rho_s^2+5)\over 32R^4\rho_s}
-{3k^2z\over 4\rho_sR^3}
-{(56k^2\rho_s^2+25)\over 64\rho_s^3R^2}+\nonumber \\
+{k^2z\over 4\rho_s^3R}(6k^2\rho_s^2-7)
-{5k^4\over 4\rho_s}+{k^2\over\rho_s^3}+{75\over 64\rho_s^5}
\Big]\cos[2k(\sqrt{z^2+\rho_s^2}-z)+\Psi_0]+\nonumber \\
+{5zx_s\over 128\rho_s^6}(128k^6\rho_s^6-160k^4\rho_s^4-336k^2\rho_s^2-105)
C_0(z,2,2\Psi_1)-\nonumber \\
-{kzx_s\over 64\rho_s^6}(64k^6\rho_s^6-320k^4\rho_s^4+980k^2\rho_s^2+525)
[S_0(z,1,2\Psi_1)-S_1(z,2,2\Psi_1)]-\nonumber \\
-{k^2x_s\over 32\rho_s^4}(176k^4\rho_s^4-176k^2\rho_s^2-75)[C_1(z,2,2\Psi_1)
-C_0(z,1,2\Psi_1)]+\nonumber \\
+{k^2\over 16\rho_s^3}(16k^4\rho_s^4+4k^2\rho_s^2-3)[C_0(z,1,\Psi)
-C_1(z,2,\Psi_0)]+\nonumber \\
+{1\over 64\rho_s^5}(64k^6\rho_s^6+32k^4\rho_s^4+276k^2\rho_s^2+75)
[k\rho_sx_sS_0(z,2,2\Psi_1)+zC_0(z,2,\Psi_0)]+\nonumber \\
+{k\over 32\rho_s^3}(80k^4\rho_s^4+3)S_0(z,2,\Psi_0)
+{kz\over 32\rho_s^5}(176k^4\rho_s^4-176k^2\rho_s^2-75)\times\nonumber \\
\times[S_1(z,2,\Psi_0)-S_0(z,1,\Psi_0)],
\end{align}
 $\Psi_0=2k(z_s-ct_s)+\varphi_s,$
$2\Psi_1=2k(z_s-ct_s)+2\varphi_s.$

Similarly, for the dependence of the coordinate $y$ of the ray on the coordinate $z$, we find:
\begin{equation}\label{eq17}
y(z)=y_s+\eta\xi{\bf m}^2\Big\{F_y(z)-F_y(z_s)+
{(z-z_s)\over (z_d-z_s)}\Big[F_y(z_s)-F_y(z_d)\Big]\Big\},
\end{equation}
Where the notation has been introduced:

\begin{eqnarray} \label{eq18}
F_y(z)=
-{9\rho_s^2y_s\over 8R^8}-{y_s(8k^2\rho_s^2-15)\over 16R^6}
-{(16k^4\rho_s^4y_s-24k^2\rho_s^2y_s-64kzx_s-57y_s)\over 64\rho_s^2R^4}
+\nonumber \\
+{4k^3x_s\over 3R^3}+{1\over 128\rho_s^4R^2}(16k^4\rho_s^4 y_s
+256k^3\rho_s^2zx_s+120k^2\rho_s^2y_s+192kzx_s+285y_s)-\nonumber \\
-{4k^4zy_s\over 3\rho_s^2R}-{1\over 128\rho_s^7}
(304k^4z\rho_s^4y_s-256k^3\rho_s^4x_s+360k^2\rho_s^2zy_s
-192k\rho_s^2x_s+\nonumber \\
+855zy_s)arctg\Big({z\over \rho_s}\Big)
-y_s\Big[{9k\rho_s^2\over 4R^7}-{k(4k^2\rho_s^2+5)\over 8R^5}
+{5kz\over 16\rho_s^2R^4}+\nonumber \\
+{k(4k^2\rho_s^2-35)\over 32\rho_s^2R^3}
+{kz(16k^2\rho_s^2+25)\over 16\rho_s^4R^2}
-{k\over 64\rho_s^4R}(16k^4\rho_s^4-68k^2\rho_s^2+175)-\nonumber \\
-{k(z+R)\over 64\rho_s^6}(144k^4\rho_s^4-64k^2\rho_s^2-525)
\Big]\sin[2k(\sqrt{z^2+\rho_s^2}-z)+2\Psi_1]-\nonumber \\
-y_s\Big[{9\rho_s^2\over 8R^8}-{(28k^2\rho_s^2+5)\over 16R^6}
-{(12k^2\rho_s^2+35)\over 64\rho_s^2R^4}
-{5k^2z\over 8\rho_s^2R^3}+\nonumber \\
+{(16k^4\rho_s^4-180k^2\rho_s^2-175)\over 128\rho_s^4R^2}
+{5k^2z\over 8\rho_s^4R}(4k^2\rho_s^2-5)+\nonumber \\
+{(64k^6\rho_s^6+580k^2\rho_s^2+525)\over 128\rho_s^6}
\Big]\cos[2k(\sqrt{z^2+\rho_s^2}-z)+2\Psi_1]+\nonumber \\
+\Big[{5k\rho_s\over 4R^5}+{3kz\over 8R^4\rho_s}
-{5k\over 16\rho_sR^3}
+{kz(6k^2\rho_s^2+7)\over 8\rho_s^3R^2}
+{k\over 32\rho_s^3R}(16k^2\rho_s^2-25)+\nonumber \\
+{k(z+R)\over 32\rho_s^5}(16k^4\rho_s^4-28k^2\rho_s^2+75)
\Big]\cos[2k(\sqrt{z^2+\rho_s^2}-z)+\Psi_0]-\nonumber \\
-\Big[{5\rho_s\over 8R^6}-{(48k^2\rho_s^2+5)\over 32R^4\rho_s}
-{3k^2z\over 4\rho_sR^3}
-{(56k^2\rho_s^2+25)\over 64\rho_s^3R^2}+\nonumber \\
+{k^2z\over 4\rho_s^3R}(6k^2\rho_s^2-7)
-{5k^4\over 4\rho_s}+{k^2\over\rho_s^3}+{75\over 64\rho_s^5}
\Big]\sin[2k(\sqrt{z^2+\rho_s^2}-z)+\Psi_0]+\nonumber \\
+{5zy_s\over 128\rho_s^6}(128k^6\rho_s^6-160k^4\rho_s^4-336k^2\rho_s^2-105)
C_0(z,2,2\Psi_1)-\nonumber \\
-{kzy_s\over 64\rho_s^6}(64k^6\rho_s^6-320k^4\rho_s^4+980k^2\rho_s^2+525)
[S_0(z,1,2\Psi_1)-S_1(z,2,2\Psi_1)]-\nonumber \\
-{k^2y_s\over 32\rho_s^4}(176k^4\rho_s^4-176k^2\rho_s^2-75)[C_1(z,2,2\Psi_1)
-C_0(z,1,2\Psi_1)]-\nonumber \\
-{k^2\over 16\rho_s^3}(16k^4\rho_s^4+4k^2\rho_s^2-3)[S_0(z,1,\Psi_0)
-S_1(z,2,\Psi_0)]+\nonumber \\
+{1\over 64\rho_s^5}(64k^6\rho_s^6+32k^4\rho_s^4+276k^2\rho_s^2+75)
[k\rho_sy_sS_0(z,2,2\Psi_1)-zS_0(z,2,\Psi_0)]+\nonumber \\
+{k\over 32\rho_s^3}(80k^4\rho_s^4+3)C_0(z,2,\Psi_0)
+{kz\over 32\rho_s^5}(176k^4\rho_s^4-176k^2\rho_s^2-75)\times\nonumber \\
\times[C_1(z,2,\Psi_0)-C_0(z,1,\Psi_0)].
\end{eqnarray}
The expressions \eqref{eq15} together with \eqref{eq16} and \eqref{eq17} together  with  \eqref{eq18} define two different rays 
in parametric form (for $\eta=\eta_1$ and for $\eta=\eta_2$), passing through 
the point ${\bf r}_s=\{x_s,y_s,z_s\},$ where at the time $t=t_s$ the X-ray and gamma radiation burst occurs, and the point 
${\bf r}_d=\{x_s,y_s,z_d\},$
where the detector of this radiation is located. The parameter in these expressions is the coordinate $z.$

\section{The law of motion of X-ray and gamma-ray pulses along the rays} \label{sec5}
Solving the equation \eqref{eq13} with the homogeneous boundary condition \break $ct_1(z_s)=0,$ 
and also considering the corresponding solution in the zero approximation, we obtain for the dependence $ct=ct(z)$:
\begin{equation} \label{eq19}
ct(z)=ct_s+z-z_s+\eta\xi{\bf m}^2\big[F(z)-F(z_s)\big],
\end{equation}
where the notation has been introduced:
\begin{align} \label{eq20}
F(z)={9z\rho_s^2\over 8R^8}+{z\over 16R^6}(8k^2\rho_s^2+5)
+{z\over 64\rho_s^2R^4}(16k^4\rho_s^4+40k^2\rho_s^2+57)
-\nonumber \\
-{2k^4\rho_s^2\over 3R^3}-{z\over 128\rho_s^4R^2}(208k^4\rho_s^4
-120k^2\rho_s^2-171)+{4k^4\over R}+{1\over 128\rho_s^5}
(304k^4\rho_s^4+\nonumber \\
+120k^2\rho_s^2+171)arctg\Big({z\over \rho_s}\Big)
+\Big[{9k\rho_s^2z\over 4R^7}-{5k\rho_s^2\over 8R^6}
-{kz(4k^2\rho_s^2-5)\over 8R^5}+\nonumber \\
+{k(16k^2\rho_s^2+5)\over 32R^4}
-{kz(4k^2\rho_s^2-25)\over 32\rho_s^2R^3}
+{k(32k^2\rho_s^2+25)\over 64\rho_s^2R^2}
+{k\over 64\rho_s^4}(16k^4\rho_s^4-28k^2\rho_s^2+\nonumber \\
+75)\big(1+{z\over R}\big)\Big]
\sin[2k(\sqrt{z^2+\rho_s^2}-z)+2\Psi_1]
+\Big[{9\rho_s^2z\over 8R^8}
-{z(28k^2\rho_s^2-5)\over 16R^6}+\nonumber \\
+{5k^2\rho_s^2\over 4R^5}+{z(12k^2\rho_s^2+25)\over 64\rho_s^2 R^4}
-{5k^2\over 16R^3}-{z\over 128\rho_s^4R^2}(16k^4\rho_s^4-76k^2\rho_s^2-75)
+\nonumber \\
+{5k^2(8k^2\rho_s^2-5)\over 32\rho_s^2 R}
\Big]\cos[2k(\sqrt{z^2+\rho_s^2}-z)+2\Psi_1]
+{C_0(z,2,2\Psi_1)\over128\rho_s^4}(64k^6\rho_s^6+\nonumber \\
+32k^4\rho_s^4+276k^2\rho_s^2
+75)+{k[S_1(z,2,2\Psi_1)-S_0(z,1,2\Psi_1)]\over 64\rho_s^4}
(176k^4\rho_s^4-\nonumber \\
-176k^2\rho_s^2-75).
\end{align}
The obtained relations \eqref{eq19}, \eqref{eq20} for $\eta=\eta_1$ allows us to investigate the motion laws of the X-ray and gamma pulse transported 
along the first ray by the first normal wave, while for $\eta=\eta_2$, it describes the motion along the second ray, carried by 
the second normal wave, which has orthogonal polarization to the polarization of the first wave.

\section{Calculation of the nonlinear electrodynamics delay of electromagnetic signals carried by different normal waves}\label{sec6}

  To calculate the delay time $T$ of the pulse carried by one normal wave relative to the pulse carried by another normal wave, 
we subtract the propagation time of the first normal wave from the source to the detector from the propagation time of the second 
normal wave along the same path. As a result, we get:
\begin{equation} \label{eq21}
T={\xi(\eta_2-\eta_1){\bf m}^2\over c}\big[F(z_d)-F(z_s)\big].
\end{equation}
Let $z_s=0$. Substituting $z_d\to\infty$ and $z_s=0$ into the expression \eqref{eq20} we will have:
\begin{align*}
\lim\limits_{z_d\to\infty}F(z_d)
={\pi\over 256\rho_s^5}(304k^4\rho_s^4+120k^2\rho_s^2+171)+
{k\sin 2(\varphi_s-kct_s)\over 32\rho_s^4}(16k^4\rho_s^4-\\
-28k^2\rho_s^2+75)
+{\lim\limits_{z_d\to\infty}C_0(z_d,2,2(\varphi_s-kct_s))\over128\rho_s^4}
(64k^6\rho_s^6+32k^4\rho_s^4+276k^2\rho_s^2+75)+\\
+\lim\limits_{z_d\to\infty}{k[S_1(z_d,2,2(\varphi_s-kct_s))-S_0(z_d,1,2(\varphi_s-kct_s))]
	\over 64\rho_s^4}(176k^4\rho_s^4-176k^2\rho_s^2-\\
-75),\\
F(0)={k\sin(2k\rho_s+2(\varphi_s-kct_s))\over 32\rho_s^4}
(8k^4\rho_s^4+18k^2\rho_s^2+35)+\\
+{k^2\cos(2k\rho_s+2(\varphi_s-kct_s))\over 32\rho_s^3}
(40k^2\rho_s^2+5)+{10k^4\over 3\rho_s}.
\end{align*}

Therefore, the expression for the time delay \eqref{eq21} in the considered case will take the form:
\begin{align*}
T={\xi(\eta_2-\eta_1){\bf m}^2\over c\rho_s^5}\Big\{
{\pi\over 256}(304k^4\rho_s^4+120k^2\rho_s^2+171)+
{k\rho_s\sin 2(\varphi_s-kct_s)\over 32}\times\\
\times(16k^4\rho_s^4-28k^2\rho_s^2+75)
+{k\rho_s\over32}(64k^6\rho_s^6+32k^4\rho_s^4
+276k^2\rho_s^2+75)I_1(2(\varphi_s-kct_s))-\\
-{10k^4\rho_s^4\over 3}-{k\rho_s\over 32}
(176k^4\rho_s^4-176k^2\rho_s^2-75)I_2(2(\varphi_s-kct_s))-\\
-{k\rho_s\sin(2k\rho_s+2(\varphi_s-kct_s))\over 32}(8k^4\rho_s^4+18k^2\rho_s^2+35)-\\
-{k^2\rho_s^2\cos(2k\rho_s+2(\varphi_s-kct_s))\over 32}
(40k^2\rho_s^2+5)\Big\},
\end{align*}
where
\begin{equation*}
I_1(2(\varphi_s-kct_s))=\int\limits_0^{2k\rho_s}{\cos(\chi+\Psi)
	\over (\chi^2+a^2)}d\chi,\ 
I_2(2(\varphi_s-kct_s))=\int\limits_0^{2k\rho_s}{\chi\sin(\chi
	+\Psi)\over (\chi^2+a^2)}d\chi.
\end{equation*}

From this expression, it follows that in Heisenberg-Euler electrodynamics, the time delay $T$ in the magnetic field of a typical 
pulsar $(|{\bf m}|/R_0^3\sim 10^{13}$ Gs)reaches $10^{-8}$ seconds, and for a magnetar, it is approximately two orders of 
magnitude larger.

\section{Conclusion}\label{sec7}
Currently, with the development of satellite observations \cite{18} that register hard electromagnetic radiation coming from Sun, 
pulsars 
and magnetars, there is an opportunity to prepare an experiment to observe the effect of nonlinear electrodynamic birefringence.
One of observable manifestations of the vacuum birefringence in pulsar or neighbourhood is related to normal
mode delay for hard X-ray and gamma- radiation pulses passing near the pulsar. Because
of the vacuum birefringence, the propagation velocity of the fast mode is greater than that of the slow mode
so it will reach to the detector earlier. So the leading part of any pulse coming from the X-ray source to the detector will be 
linearly polarized due to this mode. This part of the pulse will have a time duration T. After this time, the slow mode will reach 
the detector, and the pulse polarization state will most likely change to elliptical.
When the fast mode leaves the detector, only the signal carried by the slow mode will continue to propagate through it. Therefore, 
the trailing part of any pulse will be linearly polarized perpendicular to the polarization of the leading part of the pulse. 
This polarization state of all hard pulses coming from pulsars and magnetars can only be ensured by the nonlinearity of vacuum 
electrodynamics.
To verify this prediction of vacuum nonlinear electrodynamics, it is necessary to measure the polarization state of X-ray 
and gamma pulses coming from pulsars and magnetars. 
Polarimeters for these frequency ranges are currently unavailable, but several research 
centers  \cite{19,20,21,22,23,24,25,26,27,28} have plans to develop such devices.

%%%%%%%%%%%%%%%%

\begin{thebibliography}{50}
\bibitem{1}
M.Born, L.Infeld, {\it Foundations of the New Field Theory}, {\it Proc. Roy. Soc. Lond. Ser. A}, {\bf 144} (1934) 425-451.

\bibitem{2}
W.Heisenberg, H.Euler,{\it Folgerungen aus der Diracschen Theorie des Positrons, Z. Phys.}, {\bf 89}:11-12 (1936) 714-732.

\bibitem{3} 
P.A.Vshivtseva and M.M Denisov,{\it Mathematical modeling of electromagnetic
	wave propagation in nonlinear electrodynamics, 
	Compututional Mathematics and Mathematical Physics} {\bf 49},(2009) 2092.

\bibitem{4}
V.I.Denisov V.A.Sokolov and S.I.Svertilov
{\it Vacuum nonlinear electrodynamic polarization effects in hard emission of pulsars and
	magnetars, JCAP} {\bf 09} (2017) 004.

\bibitem{5}
D.L.Burke et al.,
{\it  Positron Production in Multiphoton Light-by-Light Scattering, Phys. Rev. Lett.} {\bf 79} (1997) 1626.

\bibitem{6}
R.N.Manchester, J.H.Taylor, {\it Pulsars} (W. H. Freeman, San Francisco,
1977).

\bibitem{7}
R.N.Manchester, G.B.Hobbs, A.Teoh, and M.Hobbs,
{\it The Australia telescope national facility pulsar catalogue,
	Astron. J.} {\bf 129}  (2005) 1993.

\bibitem{8}
S.A.Olausen and V.M.Kaspi, {\it  The McGill magnetar catalog,
	Astrophys. J. Suppl.} {\bf 212} (2014) 1.

\bibitem{9}
L.Sidoli, A.Paizis and K.Postnov, {\it INTEGRAL study of temporal properties of bright flares in
	Supergiant Fast X-ray Transients, Mon. Not. Roy. Astron. Soc.} {\bf 457} (2016) 3693.

\bibitem{10}
Jin Young Kim and Taekoon Lee, {\it  Light bending by nonlinear electrodynamics under strong electric and magnetic field, 
	Journal of Cosmology and Astroparticle Physics} {\bf 11} (2011) 017.

\bibitem{11}
Medeu Abishev et al., 
{\it Some effects of nonlinear vacuum electrodynamics in strong magnetic and gravitational fields of the pulsar, Astroparticle 
	Physics} {\bf  73} (2016) 8-13.

\bibitem{12}
M.I.Vasil'ev and V.A.Sokolov {\it Nonlinear-electrodynamic effects in the electromagnetic field of a rotating pulsar.,
	Moscow Univ. Phys.} {\bf 67} (2012) 418-422  https://doi.org/10.3103/S002713491205013X

\bibitem{13}
V.I.Denisov, I.P.Denisova and V.A.Sokolov,  {\it Using the concept of natural geometry in the nonlinear electrodynamics 
	of the vacuum, Theor Math Phys.}{\bf 172}  (2012) 1321-1327, https://doi.org/10.1007/s11232-012-0117-3

\bibitem{14}
V.I.Denisov  and  S.I.Svertilov {\it Vacuum nonlinear electrodynamics curvature 
	of photon trajectories in pulsars and magnetars, 
	Astronomy and Astrophysics,} {\bf 399},  (2003) L39-L42.

\bibitem{15}
Elda Guzman-Herrera, Ariadna Montiel and Nora Breton, 
{\it Comparative of light propagation in Born-Infeld, Euler-Heisenberg and Modmax nonlinear electrodynamics}, JCAP, 
	{\bf 11} (2024) 002.
	
	\bibitem{16}	
L.D. Landau, E.M. Lifshitz, {\it The Classical Theory of Fields},\break 
(Butterworth-Heinemann, 1975).
	
	\bibitem{17}	
I.S.Gradshteyn and I.M.Ryzhik, {\it Table of Integrals, Series, and Products} Fifth Edition,
Academic Press, UK (1994).
	
	\bibitem{18}	
A.V.Bogomolov et al., {\it Temporal, spectral and polarization parameters of hard X-ray emission of solar flares observed in SPR-N experiment on board CORONAS-F orbital observatory, Izv.
	Akad. Nauk Ser. Fiz.} {\bf 36} (2003) 1422.


\bibitem{19}	
E. Caroli et al., A CdTe position sensitive spectrometer for hard X- and soft -ray polarimetry,
Nucl. Instrum. Meth. A 477 (2002) 567.

\bibitem{20}
A. Zoglauer et al., Polarization Measurements with the MEGA Telescope, Proceedings of the 5th
INTEGRAL Workshop on the INTEGRAL Universe (ESA SP-552), Munich Germany (2004).

\bibitem{21}	
R. Bellazzini et al., A Sealed Gas Pixel Detector for X-ray Astronomy,
Nucl. Instrum. Meth. A 579 (2007) 853. 
	
\bibitem{22}	
E. Aprile, A. Curioni, K.L. Giboni, M. Kobayashi, U.G. Oberlack and S. Zhang, Compton
Imaging of MeV Gamma-Rays with the Liquid Xenon Gamma-Ray Imaging Telescope
(LXeGRIT), Nucl. Instrum. Meth. A 593 (2008) 414. 

\bibitem{23}	
P.F. Bloser et al., Calibration of the Gamma-RAy Polarimeter Experiment (GRAPE) at a
Polarized Hard X-Ray Beam, Nucl. Instrum. Meth. A 600 (2009) 424.

\bibitem{24}	
S. Orsi et al., Response of the Compton polarimeter POLAR to polarized hard X-rays,
Nucl. Instrum. Meth. A 648 (2011) 139.

\bibitem{25} J. Greiner et al., GRIPS - Gamma-Ray Imaging, Polarimetry and Spectroscopy,
Exper. Astron. 34 (2012) 551.

\bibitem{26}	
P. Soffitta, {\it XIPE: the X-ray imaging polarimetry explorer, Exper. Astron.} {\bf 36} (2013) 523.
	
 \bibitem{27}	
 T. Tanimori et al., An Electron-Tracking Compton Telescope for a Survey of the Deep Universe
 by MeV gamma-rays, Astrophys. J. 810 (2015) 28.                                                             
	
\bibitem{28}	
M.C. Weisskopf, The Imaging X-ray Polarimetry Explorer (IXPE). Space Telescopes and
Instrumentation 2016: Ultraviolet to Gamma Ray, Proc. SPIE bf 9905 (2016) 990517.	
\end{thebibliography}
\end{document}